# Chirality-induced effective magnetic field in a phthalocyanine molecule

Shinji Miwa[1,2,*], Kouta Kondou[3], Shoya Sakamoto[1], Atsuko Nihonyanagi[3], Fumito Araoka[3], YoshiChika Otani[1,2,3], and Daigo Miyajima[3]

1. *The Institute for Solid State Physics, The University of Tokyo, Kashiwa, Chiba 277-8581, Japan*
2. *Trans-scale Quantum Science Institute, The University of Tokyo, Bunkyo, Tokyo 113-0033, Japan*
3. *RIKEN, Center for Emergent Matter Science (CEMS), Wako, Saitama 351-0198, Japan*
**miwa@issp.u-tokyo.ac.jp*

Chirality in organic molecules has attracted considerable attention in the fields of chemistry, biology, and spintronics. This paper reports on perpendicular magnetization hysteresis loops of a multilayer consisting of ultrathin Fe (001), chiral phthalocyanine molecule (($P$)- or ($M$)-PbPc-DTBPh), and MgO (001). We find a chirality-dependent shift of the hysteresis loop. Unlike the previously reported bias current induced phenomena, the result shows a chirality-induced effective magnetic field in the phthalocyanine molecule in the absence of a bias current in the system. This study opens up a new direction in the emerging field of chiral molecular spintronics.

Spintronics research has been developing using relatively simple materials with the aid of an interface. A finite spin-orbit torque associated with spin-accumulation induced by the spin-Hall effect occurs at a ferromagnet-nonmagnet interface[1,2]. Moreover, perpendicular magnetic anisotropy and the corresponding electric-field-effect appear at a ferromagnet-dielectric interface[3-5]. In both cases, relatively simple materials such as Fe, Pt, and MgO have been employed. Beyond the simple conventional materials, various studies have tried to exploit the potential of complex organic molecules for spintronics[6], for example, spin-current generation in molecules[7], control of a localized spin in a single molecule magnet[8], spin-doping using molecules[9,10], and spin-charge conversion in molecular spintronics devices[11-13].

Recent experiments have shown that a self-assembled monolayer of three-dimensional chiral DNA molecules could exhibit spin-polarized conduction depending on the molecular chirality[14-19]. This phenomenon is called the chiral-induced spin selectivity (CISS), where the steady (or transient) current in the molecule induces the spin polarization in conduction electrons. Thus, it may be beneficial to include a chiral molecule in a spintronics device consisting of multilayers with an interface. In this study, we employ a basic and well-defined structure in spintronics, i.e., Fe/MgO (001), and fabricate a multilayer using a molecular beam epitaxy method in an ultrahigh vacuum, where a monomer chiral phthalocyanine (Pc) molecule is inserted at the Fe/MgO (001) interface. We find a chirality-induced effective magnetic field in the Pc molecule. In contrast to the previous studies[14-19], the chirality-induced phenomenon occurs in the absence of a bias current in the system.

We employed (*P*)-PbPc-DTBPh and (*M*)-PbPc-DTBPh as the chiral Pc molecules (see supplementary data for synthesis and characterization), which have right- and left-handed helicities, respectively, as shown in Figs. 1(a) and 1(b). Figure 1(c) shows a schematic of the film structure. Layers of MgO (001): 5 nm/V (001): 30 nm/Fe (001): 0.7 nm/(*P*)- or (*M*)-PbPc-DTBPh: 0-0.4 nm/MgO (001): 2 nm/AlO$_x$: 5 nm were deposited on a MgO(001) substrate. The MgO substrate was annealed at 800 °C for 10 min, and the V layer was annealed at 500 °C for 15 min after the deposition. All layers were deposited at room temperature. The PbPc-DTBPh layer was deposited at an evaporation rate of 0.05 Å/s in an ultrahigh vacuum ($< 5\times10^{-7}$ Pa) using an effusion cell with an $Al_2O_3$ crucible. The thickness of the Pc molecule was defined using a quartz thickness monitor, which was calibrated using atomic force microscopy. Here, one molecular layer of PbPc-DTBPh had a nominal thickness of approximately 0.6 nm. Other layers were grown at an evaporation rate of 0.1-0.3 Å/s using an electron beam deposition technique in an ultrahigh vacuum ($<3\times10^{-7}$ Pa). Each deposition process was evaluated by reflection high-energy electron diffraction (RHEED). Figure 1(d) shows RHEED images of the 0.7 nm Fe and 2 nm MgO, which are taken along the [110] azimuth of the MgO (001) substrate. We found that 2 nm MgO grows epitaxially, despite the presence of PbPc-DTBPh molecules on the Fe(001) surface. This result is similar to that of a previous study[20], where the MgO(001) layer was grown on Fe(001)/CoPc when the thickness of the CoPc was less than two molecular layers.

Figure 2(a) shows the magnetization hysteresis loops. The data were collected via the magneto-optical Kerr effect in a polar configuration; the magnetic field was applied normal to the film plane. A hysteresis loop was measured 30 times for each configuration, and the averaged data are plotted in Fig. 2. The black, red, and blue plots in Fig. 2(a) show the hysteresis loops of Fe without adsorption of the chiral Pc, that with 0.15 nm-, and that with 0.30 nm- molecule adsorption, respectively. While the chiral Pc adsorption at the Fe/MgO interface slightly modulates the coercive field (switching field) of Fe, all the Fe layers in this study are perpendicularly magnetized regardless of the thickness of adsorbed molecules. Figure 2(b) shows the plot of the Kerr rotation angle as a function of the thickness of the adsorbed molecules for the spontaneous magnetization. The data correspond to the magnitude of the Kerr rotation angle at zero magnetic field. As shown in Fig. 2(b), the Kerr rotation angle of 0.7 nm

Fe is almost independent of the thickness of adsorbed molecules. To be precise, the chiral Pc adsorption slightly reduces the Kerr rotation angle; however, the reduction is less than 3%. Figure 2(c) shows the coercive field as a function of the chiral Pc thickness. The coercive field is defined as the field where the polarity of the Kerr rotation angle changes. From Fig. 2(c), we can see that chiral Pc reduces the coercive field of Fe regardless of the chirality. This reduction can be attributed to an adsorption-induced change in energy for magnetic anisotropy and/or magnetic domain wall creation, which would be determined by an orbital hybridization[21,22] and a charge transfer[12] at an interface. Figure 2(d) shows the center field of hysteresis, defined as the average of the positive and negative coercive fields in Fig. 2(c). The blue and pink curves are guides for the eye for the chiral Pc molecules with right-handed ((*P*)-PbPc-DTBPh) and left-handed ((*M*)-PbPc-DTBPh) helicities, respectively. Note that there is a small but significant difference in the center field between the Fe hysteresis decorated with chiral Pc molecules with right- and left-handed helicities. In other words, a chirality-dependent effective magnetic field is induced in the Pc molecule.

If we assume that there is a spontaneous magnetic moment in the chiral Pc molecule (the blue arrows in Fig. 3), where the polarity is insensitive to the Fe magnetic moment (the black arrows in Fig. 3) but is chirality-dependent, the chirality-dependent center field shift in Fig. 2(d) can be explained as follows. One of the explanations is the exchange-bias effect of the Pc molecule as in a conventional ferromagnet/antiferromagnet system[23]. The other explanation is the linear magnetoelectric effect[24,25] in the Pc molecule combined with the voltage-controlled magnetic anisotropy effect[3-5] in Fe. In the case of linear magnetoelectric effect, the electric polarization in the Pc molecule changes with the magnetization direction of Fe, which is schematically shown in Figs. 3(a) and 3(b) (Figs. 3(c) and 3(d)). Here, the electric polarization from the broken inversion symmetry in the molecule and the charge transfer from Fe are not shown for brevity. The change in the electric polarization induces perpendicular magnetic anisotropy energy of Fe via the charge doping[26]/redistribution[27] at the Fe surface. Moreover, the electric polarization induced by the magnetoelectric effect might be chirality-dependent (compare Fig. 3(a) with 3(d) and Figs. 3(b) with 3(c)). This results in magnetization-direction-dependent magnetic anisotropy in the system and induces the chirality-dependent shift of the hysteresis loop, as confirmed in Fig. 2(d). For both explanations, note that the magnetic moment is not necessarily the magnetic *dipole* moment but can be the magnetic *monopole* and/or *quadrupole* moments as a conventional linear magnetoelectric material $Cr_2O_3$[28, 29], which is discussed from the symmetry argument[30].

Previously, it was reported that a self-assembled monolayer of three-dimensional chiral DNA molecules could possess spin-polarized conduction electrons depending on the molecular chirality, which is generally called the CISS. The CISS has been confirmed in terms of various phenomena, such as circular polarization of light[14], magnetoresistance effect[15,31], molecule-adsorption-induced magnetization switching[16], anomalous Hall effect[17], and separation of enantiomers by a ferromagnet[18]. In these studies[14-19], the CISS is explained in terms of the spin polarization of conduction electrons induced by the steady (or transient) current in the molecule. In contrast to these, we employed a system with a monomer chiral molecule, without a three-dimensional helical structure, directly adsorbed onto the ferromagnetic metal, which is prepared in an ultrahigh vacuum. Moreover, there is no bias current in the experimental system. On the difference in the chiral molecules, the molecule employed in this study can be adsorbed onto the Fe surface with its face up or down. In the case of a similar PbPc molecule adsorbed on Cu (111), coexisting PbPc (up) and PbPc (down) molecules were identified by atomic force microscopy[12]. However, a finite chirality-induced effect should be observed even if the orientation of the molecule is completely random[32,33]. This is because the chiral vector direction in the Fe/Pc/MgO multilayer is determined by an intrinsic chirality of the molecule (*P* or *M*), and does not depend on the molecular orientation (up or down). Apart from the difference in the chiral molecules, the absence of a bias current in this study is crucial. For instance, the adsorption-induced magnetization switching[16] can be explained by the spin polarization in the chiral molecules, resulting from a charge-transfer-induced transient electric current during the molecule adsorption. However, there is no steady or transient current after the adsorption of the molecules. Thus, the general explanation for the phenomena associated with the CISS, that is, the current-induced spin polarization of a conduction electron in a molecule cannot explain the chirality-dependent effective magnetic field observed in this study (Fig. 2(d)). The above discussion also assumes a finite spontaneous magnetic moment to explain the chirality-dependent effective magnetic field. Its polarity is insensitive to the magnetization of the neighboring ferromagnet but depends on the molecular chirality. One of the previous reports also includes experimental data where an effective magnetic field appeared without a bias current[16]. We believe that further experimental and theoretical research is indispensable on this topic.

To summarize, we fabricated a multilayer including a chiral phthalocyanine molecule (*P*)- and (*M*)-PbPc-DTBPh at the Fe/MgO(001) interface and characterized it by the magneto-optical Kerr effect. Although there was an absence of any bias current in the system, a chirality-dependent effective magnetic field is confirmed in the perpendicular magnetization hysteresis loop. The experimental results indicate the presence of a spontaneous magnetic moment in the chiral phthalocyanine molecule, where the polarity of the magnetic moment is insensitive to the magnetization of Fe but depends on the molecular chirality. This study opens up a new direction in the emerging field of chiral molecular spintronics.

We thank K. Kimura of the University of Tokyo, E. Minamitani and A. Shitade of IMS for fruitful discussions. We also thank T. Higo and S. Nakatsuji of the University of Tokyo for MOKE measurements. This work was supported in part by JSPS KAKENHI (Nos. JP18H03880) and the Spintronics Research Network of Japan (Spin-RNJ).

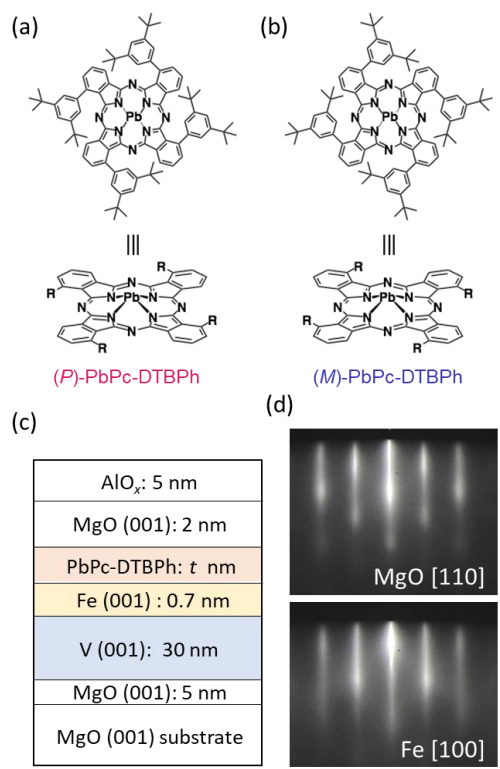

Fig. 1 (a) Structure of a chiral phthalocyanine (Pc) molecule with right-handed helicity: (*P*)-PbPc-DTBPh. (b) Structure of a chiral molecule with left-handed helicity: (*M*)-PbPc-DTBPh. (c) Schematic of the multilayer. (d) *In-situ* reflection high-energy electron diffraction images of the 2 nm MgO and 0.7 nm Fe surfaces. The incident electron beam was parallel to the [110] azimuths of the MgO substrate.

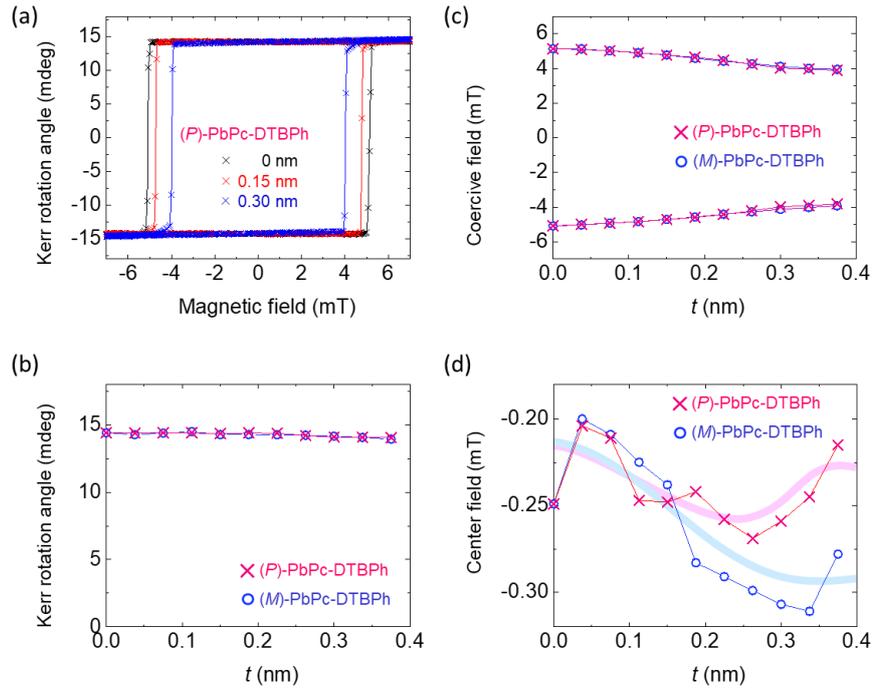

Fig. 2 (a) Magnetization hysteresis loops of Fe decorated with (*M*)-PbPc-DTBPh. A magnetic field is applied normal to the film plane. (b) Kerr rotation angle for the spontaneous magnetization as a function of PbPc-DTBPh thickness (*t*). (c) Coercive field as a function of the PbPc-DTBPh thickness (*t*). (d) Center field of the hysteresis loop, defined as the average of positive and negative coercive fields. Pink and blue curves are guides to the eye for Fe decorated with (*P*)-PbPc-DTBPh and (*M*)-PbPc-DTBPh, respectively.

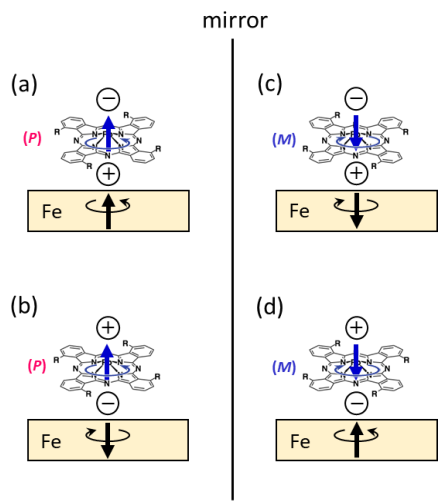

Fig. 3 (a)-(d) Schematic of possible magnetic moments in the Fe/(*P*)-((*M*)-)PbPc-DTBPh/MgO systems. Black arrows denote the magnetic moment of Fe. Blue arrows denote the chirality-dependent magnetic moment in the Pc molecules. Here, + and – denote the electric charges when we assume the linear magnetoelectric effect in the chiral Pc molecule.

# Supplementary data:
# Chirality-induced effective magnetic field in a phthalocyanine molecule

Shinji Miwa[1,2*], Kouta Kondou[3], Shoya Sakamoto[1], Atsuko Nihonyanagi[3], Fumito Araoka[3], YoshiChika Otani[1,2,3], and Daigo Miyajima[3]

1. The Institute for Solid State Physics, The University of Tokyo, Kashiwa, Chiba 277-8581, Japan
2. Trans-scale Quantum Science Institute, The University of Tokyo, Bunkyo, Tokyo 113-0033, Japan
3. RIKEN, Center for Emergent Matter Science (CEMS), Wako, Saitama 351-0198, Japan
*miwa@issp.u-tokyo.ac.jp

1. **Synthesis of PbPc-DTBPh**
2. **Optical resolution of PbPc-DTBPh**
3. **Circular dichroism spectra of (*P*)-PbPc-DTBPh and (*M*)-PbPc-DTBPh in CHCl$_3$**
4. **Reference**

---

1. **Synthesis of PbPc-DTBPh**

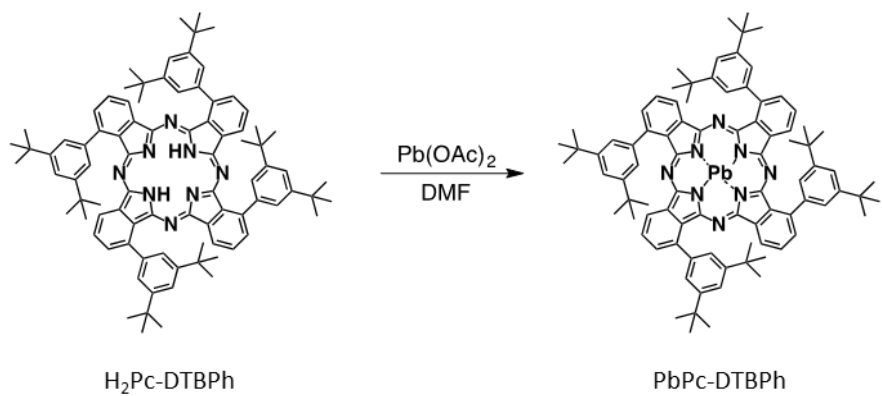

Figure S1   Synthesis of PbPc-DTBPh.

As shown in Fig. S1, H$_2$Pc-DTBPh was synthesized according to a previously reported synthetic method (Ref. S1). In a solution of H$_2$Pc-DTBPh (30 mg, 0.024 mmol) in DMF (1 mL), Pb(OAc)$_2$ (18 mg, 0.048 mmol) was added at room temperature, and the resulting solution was stirred under argon at 120 °C for 19 h. The reaction mixture was concentrated under reduced pressure, and the residue was purified by column chromatography on silica gel (20% CH$_2$Cl$_2$ in hexane as eluent). Yield: 22 mg (0.015 mmol, 62%), which was a green solid. $^1$H NMR (600MHz, CDCl3) δ 1.52 (s, 72H), 7.88-7.97 (m, 20H), 8.23 (d, J = 7.2 Hz, 4H). MS (MALDI-TOF): 1472.31 [M]$^+$

## 2. Optical resolution of PbPc-DTBPh

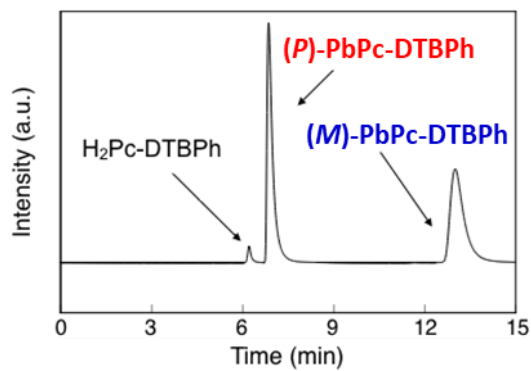

Figure S2  Optical resolution results.

Optical resolution of PbPc-DTBPh was determined using HPLC CHIRALPAK ID (DAICEL) as the chiral stationary phase and a mixed solution (Hexane: DCM = 85:15) as the eluent at a flow rate of 10 mL min$^{-1}$. As shown in Fig. 2, the peaks were monitored by absorption of 720 nm light. Note that we could see a peak corresponding to H$_2$Pc-DTBPh in this optical resolution process due to the insufficient purification of PbPc-DTBPh.

## 3. Circular dichroism spectra of (*P*)-PbPc-DTBPh and (*M*)-PbPc-DTBPh in CHCl$_3$

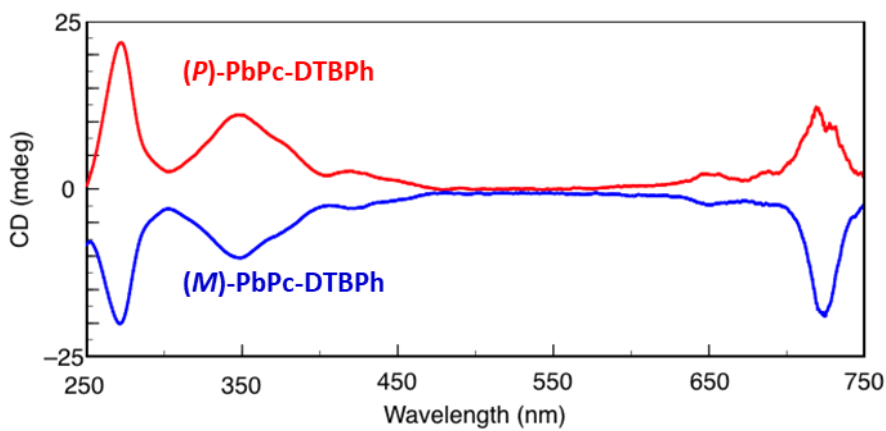

Figure S3  Circular dichroism spectra.

Figure S3 shows circular dichroism (CD) spectra of (*P*)-PbPc-DTBPh and (*M*)-PbPc-DTBPh in CHCl$_3$ after the optical resolution.